\documentclass[11pt]{JHEP3}
\usepackage{amsfonts,amsmath,amssymb,multirow}
\newcommand{\be}{\begin{equation}}
\newcommand{\ee}{\end{equation}}
\newcommand{\ba}{\begin{array}}
\newcommand{\ea}{\end{array}}

\newcommand{\bea}{\begin{eqnarray}}
\newcommand{\eea}{\end{eqnarray}}

\newcommand{\nn}{\nonumber}

\newcommand{\ra}{\rightarrow}
\newcommand{\cN}{{\cal N}}

\newcommand{\Tr}{\mbox{Tr}}

\newcommand{\cX}{{\cal X}}
\newcommand{\cC}{{\cal C}}

\newcommand{\cT}{{\cal T}}
\newcommand{\tr}{{\rm tr}}
\title{M-brane bound states and the supersymmetry of \\ BPS solutions in the Bagger-Lambert theory}
\author{Imtak Jeon, Jongwook Kim, Bum-Hoon Lee and Jeong-Hyuck Park \\
Department of Physics and Center for Quantum Spacetime,
\\
Sogang University, Seoul 121-742, Korea }
\author{Nakwoo Kim \\
Department of Physics and Research Institute of Basic Science, \\
Kyung Hee University, Seoul 130-701, Korea \\
}

\date{\today}
\preprint{}
\abstract{We continue our study of BPS equations and supersymmetric configurations in the
Bagger-Lambert theory. The superalgebra allows three different types of central
extensions which correspond to compounds of various M-theory objects: M2-branes, M5-branes,
gravity waves and Kaluza-Klein monopoles which intersect or have overlaps with the M2-branes
whose dynamics is given by the Bagger-Lambert action.
As elementary objects they are all 1/2-BPS, and multiple intersections
of $n$-branes generically break the supersymmetry into $1/2^n$, as it is well known.
But a particular composite of M-branes can preserve from 1/16 up to 3/4 of the original
${\cal N}=8$ supersymmetries as previously discovered. In this paper we provide
the M-theory interpretation for various BPS equations, and
also present explicit solutions to some 1/2-BPS equations.
}
\begin{document}

\section{Introduction}
Recently in a series of papers Bagger and Lambert (BL) have put forward a very intriguing proposal
for the action of multiple M2-branes \cite{Bagger:2006sk,Bagger:2007jr,Bagger:2007vi}.
The BL theory is a three dimensional, Chern-Simons-matter system with maximal superconformal invariance.
The action is based on a new type of gauge symmetry generated by so-called 3-algebra,
see also the independent developments by Gustavsson \cite{gust}.
For the unique realization as a conventional, ghost-free field theory with a finite number of fields,
the BL theory simply exhibits an ordinary $SO(4)\simeq SU(2)\times SU(2)$ Yang-Mills invariance with opposite
levels for the two Chern-Simons terms, and matter fields come in bi-fundamental representations.
 The uniqueness is due to the surprisingly strong restriction imposed by the (closure of) 3-algebra
 \cite{nogo}. In order to generalize one can consider Lorentzian gauge groups, see
 \cite{lorentzian} for discussions along that direction, or consider the large-$n$ limit, i.e. infinite dimensional vector space for 3-algebra,
 realized for instance as the volume-preserving diffeomorphisms of an auxiliary, three dimensional metric
 space $\Sigma$. The latter description might have a natural origin as an M5-brane action,
 see e.g. \cite{2m5}.

If the BL theory is to provide an authentic description of multiple M2-branes,
it must be able to incorporate the various M-branes which are known to exist.
They are supersymmetric objects of the 11 dimensional quantum gravity, and will
appear as classical BPS solutions in the dual field theory, the BL lagrangian.
 Simple 1/2-BPS equations can be readily written and also the solutions have been studied,
 see \cite{Bagger:2006sk,Bagger:2007vi}. A more systematic classification of BPS
  configurations have been undertaken recently by some of the present authors \cite{bps1},
  and a host of BPS equations are found with diverse numbers of unbroken supersymmetries
  ranging from 1 to 12. We however did not attempt to give a full M-theory interpretations
  or discussed new explicit solutions.

In this companion paper, we aim to fill this gap and examine the BPS equations to identify
the M-brane configurations. The terminology 'M-brane' usually means just M2 and
M5-branes in the literature, but in this paper it will be sometimes used to represent
a more general class of 'M-theory objects', e.g. including M-waves and Kaluza-Klein monopoles (KK).
Since M-theory basically still makes sense as certain limits of some string theory,
M-theory objects are sometimes more conveniently understood as originating from D-branes
or NS5-branes in IIA string theory. As it is very well known, fundamental strings and
D2-branes become M2-branes, while D4 and NS5-branes similarly merge into M5-branes, when we uplift to 11 dimensions.
D0 and D6-branes on the other hand become geometric objects: gravity waves and Kaluza-Klein monopoles.

A given BPS equation of BL theory thus depicts a certain compound of M-objects in general.
Intuitively they represent a set of different M-branes intersecting with each other.
Each element of a given set generically break the supersymmetry into one half, and the
minimal supersymmetry turns out to be one, or 1/16 of the original supersymmetry of BL action.
But it is possible that they conspire to preserve enhanced number of supersymmetries.
We will briefly explain when and how such phenomena can happen.

A remark is in order here, on the interpretation of Bagger-Lambert theory as a theory of M2-branes.
The BL theory has only one coupling constant, which is quantized due to the existence of Chern-Simons terms.
The level of the Chern-Simons term, which we call $k$, can be traced back to an orbifold action,
through discrete identifications of the fields \cite{Lambert:2008et}.
It is claimed that the BL theory with $SO(4)$ gauge group describes the motion of two
M2-branes in an orbifold $\mathbb{R}^8/D_{2k}$ in general. ($D_{2k}$ is the dihedral group.)

It is obvious that the classical solutions we will present in this
paper must be compatible with the orbifold action. As usual, $k$
can be absorbed into the re-scaling of the fields and the
classical equation of motion is independent of $k$. As a matter of
convenience, in this paper we will not put too much emphasis on
different values of $k$. We will just provide the M-brane
interpretation in {\it flat} background (i.e. $k=1$), although it
is always understood that the BL theory retains only the degrees
of freedom which are invariant under the orbifold projection.

It is natural to classify the BPS equations into three categories. Vortices with cohomogeneity two,
domain walls with cohomogeneity one, or the spacetime-filling configurations.
It will become evident that they are basically M2-branes, M5-branes and Kaluza-Klein monopoles
 which intersect with,
or completely embrace, the M2-branes responsible for the BL action we started with.

We also study explicit solutions to some 1/2-BPS equations. In
particular, with certain simplifying assumptions the 1/2-BPS
vortex equations are reduced to the Liouville and sinh-Gordon
equations, albeit with wrong-sign potential terms. We present a
few nontrivial solutions to them.

This paper is organized as follows. In Section 2 we briefly review the supersmmetric configurations
of intersecting M-branes. We also discuss the generalization to non-commutative M-branes which will
be relevant to some BPS configurations. In Section 3 we consider the superalgebra of Bagger-Lambert
theory and identify each central terms as a certain composite of M-branes.
In Section 4 we study the three classes of BPS equations, give M-theory interpretations and discuss
the preserved number of supersymmetries for some exotic cases. We
conclude with discussions in the final section.

\section{M-theory objects and their intersections}
According to the standard understanding M-theory dynamics is given in terms of gravitons,
a 3-form gauge field, and the superpartner gravitinos. There do exist objects charged with
respect to the gauge field: they are extended objects, and the electric ones are called M2-branes
while the magnetic ones are M5-branes. As a matter of fact purely gravitational,
or geometrical objects also exist as a pair. It is sometimes advantageous to view
them as the dimensional uplifts of D0 and D6-branes in IIA string theory.
The charge of D0-branes gets mapped to the momentum of gravitational waves along the so-called
11th direction. D6-branes become the Kaluza-Klein monopoles (KK) in 11 dimensions.
A KK is a 4-dimensional configuration, so in 11 dimensions one is left with 6+1 dimensional
Lorentz invariance, and it is why a KK is sometimes called M6-branes in the literature.

Like their IIA counterparts, the four types of 'M-branes' reminded above are all 1/2-BPS
configurations on their own. Strictly speaking however, M-branes have their own dynamics,
and an M-brane preserves 1/2 of the supersymmetries when it is at the vacuum configuration.
In string theory, worldvolume gauge field couples to the (pull-back of) background Ramond-Ramond
fields, and in particular a non-zero magnetic field implies that the D$p$-brane in fact has formed
a bound state
with D$(p-2)$-branes. Analogously in M-theory, if an M5-brane has a nontrivial gauge field,
it is a result of M2-branes dissolved into the M5-brane worldvolume.

The precise requirement of preserved supersymmetry for
intersecting D-branes or M-branes, with or without worldvolume
gauge field excitations, had been extensively studied in 90's. In
this section we will give a review at an elementary level to
recapitulate what is needed for our discussions in this paper. For
those readers who need more detailed information on intersecting
M-branes, especially as supergravity solutions, references in
\cite{sugra_review} may provide a decent starting point. For the
supersymmetry projection rules for intersecting brane
configurations, see e.g. \cite{proj_rule}.

Let us first consider intersections of M-branes at the vacuum configuration, i.e. without field excitations.
A compound of M-branes remains supersymmetric only when they are arranged in specific ways. A sufficient condition can be summarized as follows: a set of M-branes are supersymmetric if every possible pairing of different M-branes in the set satisfy the so-called intersection rule.

The intersection rule basically derives from the compatibility condition of Killing spinor projection rules.
For M-branes, it turns out that the 1/2-BPS projection rules are given in Table~\ref{m:projection},
where the numeric indices of the gamma matrices denote the part of 11 dimensional spacetime along
which the M-brane is extended, and $\epsilon$ is an 11 dimensional Majorana spinor.
\begin{table}[htb]
\begin{center}
\begin{tabular}{|c|c|c|c|c|}
  \hline
  M-branes & MW & M2 & M5 & KK \\
  \hline
  Projection rules & $\Gamma_{01}\epsilon=\epsilon$  & $\Gamma_{012}\epsilon=\epsilon$ & $\Gamma_{012345}\epsilon=\epsilon$ & $\Gamma_{0123456}\epsilon=\epsilon$ \\
  \hline
\end{tabular}
\end{center}
\caption{BPS projection rules for 1/2-BPS M-branes}
\label{m:projection}
\end{table}

Compatibility implies that the associated projectors such as $P=1\pm\Gamma_{012}$ for
an M2-brane should commute with each other for each pair of different M-branes.
It is then a very simple matter to verify the following intersection rules for 1/4-BPS configurations,
given in Table~\ref{m:intersection}.
In the table the numbers denote the dimensionality of M-brane worldvolume shared
by the two M-branes in question.
It is trivial that two parallel M-branes standing in parallel preserve
the same supersymmetry so such configurations are not explicitly presented in
Table~\ref{m:intersection}.

\begin{table}[htb]
\begin{center}
    \begin{tabular}{|c|c|c|c|c|}
       \hline
         & MW & M2 & M5 & KK \\
         \hline
       MW & N/A & 1 & 1 & 1 \\
       M2 & - & 0 & 1 & 0,2 \\
       M5 & - & - & 1,3 & 3,5 \\
       KK & - & - & - & 2,4 \\
       \hline
     \end{tabular}
\end{center}
\label{m:intersection}
\caption{Intersection rules for M-branes}
\end{table}

For instance, a pair of M2-branes can intersect over a point to constitute a 1/4-BPS configuration.
We will use a self-explanatory notation and represent it as (M2,M2$|0$).
One can give more detailed information in a table: for instance the following table gives a particular
(M2,M5$|1$) configuration:
\begin{center}
\begin{tabular}{cccccccccccc}
   M2: & 0 & 1 & 2 & - & - & -& - & - & - & - & - \\
   M5: & 0 & 1 & - & 3 & 4 & 5 & 6 & - & - & - & - \\
\end{tabular}
\end{center}
where an M2-brane is extended along 1,2 directions, while an M5-brane is put along 1,3,4,5,6.
They share a one dimensional subspace, satisfy the intersection rule and preserve 1/4 of the total 32
supersymmetries in general.

Let us give another example which involve more branes:
\begin{center}
    \begin{tabular}{cccccccccccc}
       M5: & 1 & 2 & 3 & 4 & 5 & - & - & - & - & - & - \\
       M5: & 1 & 2 & 3 & - & - & 6 & 7 & - & - & - & - \\
       M2: & - & - & - & 4 & - & 6 & - & - & - & - & - \\
       M2: & - & - & - & - & 5 & - & 7 & - & - & - & - \\
     \end{tabular}
\end{center}
which is 1/8-BPS.
One can easily check that every single pair from the set conform to the intersection rule.

In the examples given above all the M-branes are supposed to be at the vacuum configuration, i.e. without gauge field excitations in particular. It is in fact possible to turn on the worldvolume gauge fields without breaking supersymmetry. A typical example is a D0-D4 bound state in IIA string theory, which is 1/4-BPS. It is possible to turn on magnetic fields on the D4-brane, and then the D4-brane worldvolume configuration becomes non-commutative instantons. As D-branes this can be described as a D0-D2-D2-D4 bound state which is still 1/4-BPS in general. The M-theory lift gives (MW,M5$|1$) configuration which is 1/4-BPS, and with worldvolume fields, it turns into a MW-M2-M2-M5 bound state. One can determine the field strength or the M2-brane charges required to guarantee unbroken supersymmetries by considering the BPS projection rule:
\be
(p \Gamma_{05} +\xi_1 \Gamma_{012} + \xi_2 \Gamma_{034} +  y \Gamma_{012345} )\epsilon
=\epsilon
\ee
where the coefficients $\xi_1,\xi_2,p,y$ are related to charges of MW(5), M2(12), M2(34), and M5(12345)-branes respectively. One can easily verify that the above eigenvalue problem allows eight linearly independent solutions, i.e. becomes 1/4-BPS if
\be
(y\pm p)^2 + (\xi_1 \mp \xi_2)^2 = 1
\label{ncpro}
\ee
For more details and on the construction of supergravity solutions, see \cite{Bergshoeff:2000qn}.

Another example which will be also realized as BPS solutions of BL theory is a M2-M5-M5-KK bound state.
In IIA theory, it is D2-D4-D4-D6 bound state, which is related to D0-D2-D2-D4 solution through simple T-dualities.
The supersymmetry projection is given as
\be
(p' \Gamma_{012} +\xi'_1 \Gamma_{012347} + \xi'_2 \Gamma_{012567} +  y' \Gamma_{0123456} )\epsilon
=\epsilon
\ee
and the 1/4-BPS condition is written identically to Eq.(\ref{ncpro}), now in terms of the primed symbols,
$p',\xi'_1,\xi'_2,y'$.

\section{Superalgebra of Bagger-Lambert theory and M-branes}
\subsection{The Bagger-Lambert action and representations of 3-algebra}
The spectrum of supersymmetric solitons in M2-brane theory and their intersection
rules can be seen encoded in the worldvolume supersymmetry algebra.
The central extension of the M2-brane worldvolume supersymmetry has been derived in
\cite{Bergshoeff:1997bh}, which contains new $p$-form charges for intersecting M2 and M5-branes.
As a proposal for multiple M2-brane action, the superalgebra of BL theory should reproduce the
result in \cite{Bergshoeff:1997bh}. This task has been performed first in \cite{Passerini:2008qt}
and a further investigation can be found in \cite{Furuuchi:2008ki}.

Let us here briefly introduce the Bagger-Lambert theory, mainly in order to setup the notation.
 The BL action is given as follows:
\bea
{\cal L} &=&
-\frac{1}{2} D_\mu X^{aI} D^{\mu} X^I_a
+
\frac{i}{2} \bar{\Psi}^a \Gamma^\mu D_\mu \Psi_a
+
\frac{i}{4} \bar{\Psi}_b\Gamma_{IJ} X^I_c X^J_d \Psi_a f^{abcd}
\nn\\
&&
- V
+
\frac{1}{2} \varepsilon^{\mu\nu\lambda}
(f^{abcd} A_{\mu ab} \partial_\nu A_{\lambda cd} +
\frac{2}{3} f^{cda}_{\;\;\;\;\;\; g} f^{efgb} A_{\mu ab}
A_{\nu cd} A_{\lambda e f}
)
.
\eea
where $X^I_a,I=1,2,\cdots,8$ is the scalar field giving the positions of the M2-branes
in the transverse space $\mathbb{R}^8$, while $\Psi_a$ is the 16-component,
superpartner fermion with chirality condition $\Gamma_{0xy}\Psi=-\Psi$. $A_{\mu ab}, (\mu=0,x,y)$ is
the 2+1 dimensional nonabelian vector field. The covariant derivative is defined as follows:
\be
D_\mu X_a = \partial_\mu - \tilde{A}_{\mu\; a}^{\;\; b} X_b
\ee
in terms of the 'dual' gauge field $\tilde{A}_{\mu ab}={\textstyle\frac{1}{2}}f_{ab}^{\;\;cd}A_{\mu cd}$.
Since the gauge field has only  Chern-Simons-like first-order derivative term and without the ordinary Maxwell-like kinetic term,
it does not lead to any propagating degrees of freedom.

The matter fields take values in a vector space which are endowed with a 3-product structure.
The fields are expanded in terms of basis vectors, for instance $X^I = X^I_a T^a$.
The structure constants $f^{abcd}$ are totally anisymmetric and defined by
\be
[T^a , T^b , T^c ] = f^{abc}_{\;\;\;\;\;\; d} T^d
\ee
We assume there is a notion of metric $h^{ab}$ in the 3-algebra space so that one can raise or
lower the gauge indices.

$V$ is the scalar potential given as
\be
V(X) = \frac{1}{12} \sum_{I,J,K} \Tr \left( [X^I,X^J,X^K][X^I,X^J,X^K]\right)
\ee
where $\Tr (T^a T^b) = h^{ab}$.

The above action is invariant under the following supersymmetry transformation rules (see \cite{Bagger:2007jr}),
\bea
\delta X^I_a
&=&
i \bar{\epsilon} \Gamma^I \Psi_a
\\
\delta \Psi_a
&=&
D_\mu X^I_a \Gamma^\mu \Gamma^I \epsilon
 - \frac{1}{6} X^I_b X^J_c X^K_d f^{bcd}_{\;\;\;\;\;\; a} \Gamma^{IJK} \epsilon
\\
\delta \tilde{A}^{\;\;b}_{\mu\;\; a} &=&
i \bar{\epsilon} \Gamma_\mu \Gamma_I X^I_c \Psi_d f^{cdb}_{\;\;\;\;\;\; a}
\eea

In order to establish the supersymmetry one needs to impose a couple of consistency
conditions on the 3-algebra. The following is analogous to the cyclicity of ordinary trace operation,
and called {\it invariance identity}:
\bea
\Tr ( [ T^a , T^b , T^c ] , T^d ) = -
\Tr ( T^a , [ T^b , T^c , T^d ]  )
\eea
And the closure of successive supersymmetry transformations up to gauge transformation
requires the following identity, which is analogous to the Jacobi identity and dubbed {\it fundamental identity}:
\bea
[T^a, T^b , [ T^c , T^d , T^e ]]
&=&
[[ T^a , T^b , T^c ] , T^d , T^e]
+
[T^c , [ T^a , T^b , T^d ] , T^e ]
\nn\\
&+&
[ T^c , T^d , [T^a , T^b , T^e ]] .
\eea

If one further demands finite-dimensionality and positive-definite norm for the vector space $\{ T^a \}$,
there is only one nontrivial example \cite{nogo}: $SO(4)$.
In that case $a=1,2,3,4$ and one can set $f_{abcd}=\pi\varepsilon_{abcd}/k$,
where $k$ plays a role of coupling constant.
In quantum theory $k$ has to be integral in order to have a well-defined path integral, as pointed out in
\cite{Bagger:2007vi}.
In this case the BL theory becomes an ordinary $SU(2)\times SU(2)$ Yang-Mills gauge theory
with Chern-Simons action. The 'matter fields' $X,\Psi$ are in bi-fundamental representations, i.e. $(2,2)$.

It is possible to remove $k$ from the classical equation of motion. One re-scales the fields and the coupling constant as
\bea
X^I_a &\ra & \lambda^{1/2} X^I_a
\\
\Psi_a  &\ra & \lambda^{1/2} \Psi_a
\\
A_{\mu ab} &\ra & \lambda \, A_{\mu ab}
\\
f_{abcd} &\ra &\lambda^{-1} f_{abcd}
\eea
with $\lambda=k/\pi$. Then we simply have $f_{abcd}=\varepsilon_{abcd}$, and $k/\pi$ in front of the entire action.

As a proposal for alternative realization of 3-algebra, one can consider infinite-dimensional vector spaces. A natural and also intuitive example is given as the set of differentiable functions ${\cal C}^{\infty}(\Sigma)$ on a three dimensional space $\Sigma$ with metric $g$, as proposed in \cite{Bagger:2007vi}. The 3-algebra is defined as the Nambu-Poisson bracket,
\be
[ f, g, h ] = \frac{1}{\sqrt{g}} \varepsilon^{ijk} \partial_i f\partial_j g\partial_k h
\label{np}
\ee
which satisfies the invariance and the fundamental identities.

If one takes $\Sigma=S^3$ then a convenient basis is given as spherical harmonics.
More concretely, one can embed in $\mathbb{R}^4$ and consider the Cartesian coordinates
of the points on $S^3$:  $\cX^A, (A=1,2,3,4)$ satisfying $\sum (\cX^A)^2 = 1$ and give a 3-algebra structure as follows
\be
[ \cX^A, \cX^B , \cX^C ] = \varepsilon^{ABCD} \cX^D
\label{sh1}
\ee
Higher spherical harmonics are given as symmetric and traceless tensors in $\mathbb{R}^4$,
and their 3-algebras can be computed using the chain rule and Eq.(\ref{np}).

One might also simply choose $\Sigma=\mathbb{R}^3$. If we use $(X,Y,Z)$ as the Cartesian coordinates,
the basic 3-algebra relation is
\be
[ X, Y, Z ] = 1
\ee
But since $\Sigma=\mathbb{R}^3$ is now non-compact, the basis vectors are non-normalizable in this representation.
As the result, the BPS solutions we can find using this particular representation will
have an infinite energy in general. We will not reject such solutions as non-physical.
Our interpretation is that this divergence simply reflects the infinity of the total volume for
the extended objects such as M5-branes, and the finite integrand will be interpreted as
the tension of the M-branes.

In fact, one might want to envisage $\Sigma$ as part of an M5-brane worldvolume \cite{2m5}. The field theory at hand, with an infinite number of degrees of freedom and being able to describe higher-dimensional objects, has a semblance to the Matrix theory conjecture of M-theory \cite{Banks:1996vh}. It is not clear what is then the right choice for $\Sigma$, but it will be interesting to explore infinite dimensional 3-algebra and the associated Bagger-Lambert theory more seriously.

Alternatively, one might relax the condition of positive-definiteness of the norm and consider a Lorentzian three-algebra \cite{lorentzian}. In the simplest proposal one can incorporate any ordinary Lie-algebra at the price of introducing two null generators. Although such models necessarily contain ghost fields at the classical level, it is claimed that they are in fact unitary. But in the process of eliminating the negative-norm states the models are reduced to $\cN=8$ Yang-Mills model for D2-branes, so it seems unclear whether one can extract a sensible quantum field theory from a Lorentzian version of Bagger-Lambert 3-algebra theory which is genuinely different from 2+1 dimensional ordinary super-Yang-Mills theory. In this paper we thus choose not to discuss BPS solutions which are exclusively relevant to the Lorentzian 3-algebra version.

\subsection{Superalgebra, central charges and their M-brane interpretations}
The supercharges can be simply read off from the supersymmetry transformation rules,
and they are formally given as the integration of
the supercharge density \cite{Furuuchi:2008ki},
\be
Q = \int d^2x
\left(
- \Gamma^\mu \Gamma^I \Gamma^0 D_\mu X^{Ia} \Psi_a - \frac{1}{6} \Gamma^{IJK} \Gamma^0
X^I_a X^J_b X^K_c \Psi_d f^{abcd}
\right)
\ee

When we compute the anti-commutator of two supercharges, we obtain the following result
(see \cite{Passerini:2008qt,Furuuchi:2008ki}).
\bea
\{ Q^\alpha , Q^\beta \}
&=&
- 2 P_\mu (\Gamma^\mu \Gamma^0 )^{\alpha\beta} + Z_{IJ}
 (\Gamma^{IJ} \Gamma^0)^{\alpha\beta}
+ Z_{iIJKL}
(\Gamma^{IJKL}\Gamma^i\Gamma^0)^{\alpha\beta}
\nn\\&&
+ Z_{IJKL} (\Gamma^{IJKL})^{\alpha\beta}
\label{ce}
\eea
where $\alpha,\beta$ are the 11 dimensional spinor index, and $i=x,y$.

In the above we have, in addition to the usual energy momentum vector $P_\mu$, three types of central charges:
\bea
Z_{IJ} &=& - \int d^2 x \Tr ( D_i X^I D_j X^{J}\varepsilon^{ij} - D_0 X^K F^{KIJ} )
\\
Z_{iIJKL} &=& \frac{1}{3} \int d^2x \Tr
( D_j X^{[I} F^{JKL]}  \varepsilon^{ij})
\\
Z_{IJKL} &=& \frac{1}{4} \int d^2x \Tr (F^{M[IJ}F^{\;\;\;\;KL]}_{M})
\eea\label{centralcharge}
where we have also introduced a short-hand notation for 3-products: $F^{IJK}\equiv [X^I,X^J,X^K]$.
The first two classes are actually topological terms, since they can be expressed as surface integrals.

The last one, $Z_{IJKL}=Z_{[IJKL]}$ can be actually shown to vanish as well, but for a different reason:
one should make use of the invariance, the fundamental identity, and skew-symmetry.
The central charge $Z_{IJKL}$ is therefore identically zero, unless the nonabelian fields
are infinite-dimensional and have an infinite-norm.
In fact this type of central charges have been omitted in the analysis of \cite{Passerini:2008qt},
probably for this reason, and re-discovered later in \cite{Furuuchi:2008ki}: the authors
called such elements {\it non-trace}.

Configurations with non-trace elements are familiar in the Matrix theory conjecture
for M-theory in the light-cone quantization \cite{Banks:1996vh}. The large-$n$ limit of supesymmetric
Yang-Mills quantum mechanics is proposed as a non-perturbative description of M-theory, and M2-brane
solutions are described as infinite-sized matrices such as elements of Heisenberg algebra, e.g. $P,Q$,
with $[P,Q]=i$. Although they do not have a finite energy and strictly speaking one cannot posit them
as {\it solitons}, we are in the same spirit as \cite{Furuuchi:2008ki} and include such configurations
in the following discussions. It is because what we intend to describe is solitonic but extended
objects in 11 dimensions,
and by working with BL theory we are just seeing what they look like, when projected on the worldvolume of
multiple M2-branes.


Now we are almost ready to consider simple BPS equations and identify the central charge terms
as different combinations of M-branes. For BPS configurations, the right hand side of Eq.(\ref{ce}),
seen as a matrix with indices $\alpha,\beta$ becomes singular.
If we have a static configuration with $P_\mu=(E,\vec{0})$, this happens when the sum of the central
charge terms as a matrix allow an eigenvalue $E$. The degeneracy then gives the number of preserved supersymmetries.
Here
we choose not to include the conformal supersymmetry into the counting of BPS supersymmetries,
so the trivial vacuum has 16 supersymmetries. Nontrivial BPS solutions would have less supersymmetries.
They imply the existence of M-branes in addition to the 'background M2-branes' whose dynamics is the
BL theory in question.

For the first class of central charges, denoted as $Z_{IJ}$,
it is easily seen that the simplest BPS configuration is given as (anti)holomorphic curves.
For example, if we turn off all other fields except $X^1,X^2$,
\be
\partial_z X_w = 0 : \quad X_w \equiv\textstyle{\frac{1}{\sqrt{2}}} (X^1 - i X^2) , \,\,\, z \equiv \textstyle{\frac{1}{\sqrt{2}}}(x+ i y)
\label{m2m2}
\ee
and the 1/2-BPS projection rule for the eigenspinors is $(\Gamma^{xy12}\pm 1)\epsilon=0$.
Since we have two non-vanishing scalar fields as functions of worldvolume coordinates $x,y$
this is naturally seen as M2-branes intersecting with the background M2-branes.
One can also give the information as a table:
\begin{center}
\begin{tabular}{cccccccccccc}
   M2: & t & x & y & - & - & -& - & - & - & - & - \\
   M2: & t & - & - & 1 & 2 & - & - & - & - & - & - \\
\end{tabular}
\end{center}

For the next class, denoted as $Z_{iIJKL}$, we can devise as the simplest realization a configuration with four non-trivial scalar fields as functions of $y$:
\be
\partial_y X^1 \pm [ X^2 , X^3 , X^4 ] = 0 , \quad\mbox{and cyclic permutations.}
\label{gen:nahm}
\ee
In fact this is the generalized Nahm equation describing the phenomenon of the Myers effect \cite{Myers:1999ps}
where M2-branes puff up into an M5-brane, as suggested by \cite{Basu:2004ed}, and further explored especially in
relation to M-theory in \cite{g_nahm}.

More concretely if we set $X^A = f(y) \cX^A \;(A=1,2,3,4),$ where $\cX^A$ denote the basis vectors of $SO(4)$ or the $S^3$ spherical harmonics introduced in Eq.(\ref{sh1}), we can easily solve the BPS equation:
\be
f=\frac{1}{\sqrt{c\pm 2y}}
\ee
where $c$ is integration constant. In this solution the radius of $S^3$ becomes infinite at $y=\mp c/2$ so
the M2-branes puff up until they end on the worldvolume of a M5-brane,
extended along $x$ and $X^A$ directions. We thus reach the conclusion:
the central charge $Z_{iI_1I_2I_3I_4}$ describes M5-branes intersecting along $x^i$,
and extended in the transverse space along $X^{I_i}$ directions.

Now let us turn to the third class, $Z_{IJKL}$.
In the last paragraph we attained the insight that having some nonzero
'field strength' $F_{IJK}$ should be related to M5-branes.
For $Z_{IJKL}$ the relevant BPS equations do not necessarily involve spacetime derivatives,
but it is crucial to have at least two independent nonvanishing components for $F_{IJK}$.

As a simple 1/2-BPS configuration we can consider the following equation
(We turn on $X^1,\cdots, X^5$ and turn off all other fields.)
\be
F_{125} \pm F_{345} = 0 ,
\label{inin}
\ee
and the associated projection rule for the spinor $(\Gamma_{1234}\pm 1)\epsilon=0$.
One might guess that this configuration incorporates two M5-branes, one along 1,2,5 directions
and the other along 3,4,5 directions, and of course totally embracing the M2-brane worldvolume.
At first sight it looks puzzling, because although the two M5-branes with each other satisfy the
projection rule, sharing a 3 dimensional subspace $x,y,5$,
against the M2-brane they do not conform to the simple intersection rule,
which demands an M2-M5 pair should intersect over a line to be BPS.
The correct interpretation thus should include a KK, along $1,2,3,4$ directions as well as $x,y$.
This is precisely the configuration we discussed in the last paragraph of Section 2,
which is the M-theory lift of D2-D4-D4-D6 system. As a table, the BPS equation Eq.~(\ref{inin})
can be given as
\begin{table}[htb]
\begin{center}
\begin{tabular}{cccccccccccc}
   M2: & t & x & y & - & - & -& - & - & - & - & - \\
   M5: & t & x & y & 1 & 2 & - & - & 5 & - & - & - \\
   M5: & t & x & y & 1 & 2 & - & - & 5 & - & - & - \\
   KK: & t & x & y & 1 & 2 & 3 & 4 & - & - & - & - \\
\end{tabular}
\caption{Configuration of M2-branes as a noncommutative instanton in Kaluza-Klein monopole}
\label{kkinstanton}
\end{center}
\end{table}

\section{M-branes as BPS configurations of Bagger-Lambert theory}
In this section we discuss the BPS equations classified in \cite{bps1}. When necessary, we use the symbol
$N$ to denote the number of preserved supersymmetries for a given configuration. For instance,
$N=8$ means 1/2-BPS solutions of Bagger-Lambert theory which has 16 supersymmetries by construction.
\subsection{M2-branes and vortices}
Let us first consider the vortex solutions of Bagger-Lambert theory. They in general can be
described as holomorphic curves, and the simplest BPS equation is given Eq.~(\ref{m2m2}).
The associated projection condition is
\be
P \epsilon_0 = 0 : \quad P = 1 \pm \Gamma_{xy12}
\label{bps1:projection}
\ee

In fact it is possible to add extra fields to this configuration without breaking further
supersymmetry. The 1/2-BPS equation compatible with Eq.~(\ref{bps1:projection}) is
given as follows (See Eq.(3.24) of \cite{bps1})
\be
\ba{llll}
D_z X_{\bar{\omega}}=0\,,~~&~~~D_z X_{p}=0\,,~~&~~~
D_t X_{I}-i F_{I\omega {\bar \omega}} = 0\,,~~&~~~F_{I pq}=0\,,
\ea\label{bps1:eq}
\ee
where $I=1,2,\cdots,8$, $~p,q=3,4,5,6,7,8$, and
\be
\ba{ll}
X_{{\omega}}:= \textstyle{\frac{1}{\sqrt{2}}}(X_{1}-i X_{2})\,, ~&~ D_z :=\textstyle{\frac{1}{\sqrt{2}}} (D_x -iD_{y})\,.
\ea
\ee

We note that in general
we have a time-dependent configuration, when $D_t X^I\neq 0$. This obviously implies a nonzero momentum
along $X^I$ direction. From the discussions on central charges,
we now have the understanding that $F_{IJK}\neq0$ implies the existence of M5-branes.
Suppose $F_{123}$ is the only nonvanishing 3-product, and accordingly $D_t X^3\neq 0$.
Then as M-theory interpretation of Eq.~(\ref{bps1:eq}) we provide the following table:
\begin{table}[htb]
\begin{center}
\begin{tabular}{lccccccccccc}
   M2: & t & x & y & - & - & - & - & - & - & - & - \\
   M2: & t & - & - & 1 & 2 & - & - & - & - & - & - \\
   MW: & t & - & - & - & - & 3 & - & - & - & - & - \\
   M5: & t & x & y & 1 & 2 & 3 & - & - & - & - & - \\
\end{tabular}
\caption{M-theory configuration for MW-M2-M2-M5 bound state}
\label{m2:m2:kk:mw}
\end{center}
\end{table}

We can obtain other BPS equations with less supersymmetry as we turn on more 3-products and momenta.
Intuitively the corresponding M-theory interpretation is given as intersecting M2-branes, where
any possible pair from them can further expand in M5-branes with a gravity wave.

Further projection rules like $P'=1\pm \Gamma_{xy34}$ combine with $P=1\pm \Gamma_{xy12}$ and turn
into projections purely in $\mathbb{R}^8$, like $P''=1\pm\Gamma_{1234}$. According to the number of remaining supersymmetries we have
different numbers of complex structures introduced to $\mathbb{R}^8$ \cite{bps1}, and
the most general BPS configurations allowed are succinctly expressed in terms of the complex structures.
In order to describe the BPS equations with minimal supersymmetry in this class,
we need to introduce a $SU(4)$-structure in $\mathbb{R}^4$. Let us adopt $w,\zeta=1,2,3,4$ for $SU(4)$
indices. Then the most general BPS equations are given as
follows (See Eq.~({3.19}) of \cite{bps1}.)
\be
D_z X_{\bar{w}} = 0 , \quad
D_t X_w - i F_{w\zeta\bar{\zeta}} = 0 , \quad
F_{\zeta_1\zeta_2\zeta_3} = 0
\label{ve}
\ee
The first equation says the complexified scalar fields $\frac{1}{\sqrt{2}}(X^{2A-1}+i X^{2A})$ are covariantly
holomorphic. And the second demands the M-wave charge is the same as M5-brane charge, which is given
as a symplectic trace over two indices of the 3-product. The last condition implies the 3-form
tensor given by $F_{IJK}$ has no $(3,0)$ part.
For BPS equations with a different number of supersymmetries and the projection rules
 readers are referred to \cite{bps1}.

\subsubsection{Vortex Solutions}
In this subsection we employ a particular ansatz to find exact solutions to Eq.\eqref{ve}. If we have a nontrivial gauge field configuration, the BPS equations alone do not lead to the full equation of motion. One needs to impose the gauge field equation, especially the Gauss' law as an independent condition, in addition to Eq.\eqref{ve}.
\be\ba{l}
\tilde{F}_{z\bar z}{}^{a}{}_{b} + i X^I_c D_t X^I_d f^{cda}_{~~~~~b} = 0\, .
\ea\label{gausslaw}
\ee

More explicitly we choose the simplest version of BL theory with $SO(4)$ gauge group. The coupling constant $f_{abcd}$ is absorbed into the re-scaling of the fields so that the classical equations of motion do not contain the level $k$. In other words, we have $f_{abcd}=\varepsilon_{abcd}$.

We are interested in a purely bosonic configuration, whose nontrivial fields include $X^I, I=1,2,3$ and $A_\mu$. All other fields are set to zero. $X^I$ can be treated as a 4-vector, and in our ansatz
\be\ba{llll}
X_\omega \equiv {\textstyle\frac{1}{2}}(X^1 - i X^2) = \left(f_1,f_2,0,0 \right) \,,&
X_3= \left(0,0,0,v\right)\, .
\ea\ee
We are here interested in the solutions with non-vanishing M-wave and M5-brane charges, so it is assumed $v\neq 0$.
For the gauge field the only nonvanishing components are
\be
\tilde{A}_{\mu 12} = a_\mu , \quad \tilde{A}_{\mu 34} = b_\mu .
\ee
We assume all the unknown functions $f_1,f_2,v,a_\mu,b_\mu$ are independent of time $t$.

The first equation in Eq.\eqref{ve}, i.e. covariant holomorphicity, can be integrated to give
\begin{equation}
f^{2}_1+f^{2}_2=\alpha(z) ,
\label{alpha_def}
\end{equation}
where $\alpha$ is an arbitrary holomorphic function of $z$.

It would be certainly very nice if one can completely solve the remaining equations for arbitrary $\alpha$. For simplicity however, in this paper we will only consider the case of constant $\alpha$. Then without losing generality we assume $\alpha$ is a real number. We have two physically distinct classes, one with $\alpha=0$ and the other case of $\alpha\neq 0$.

Let us first consider $\alpha=0$ case. If one explicitly solves the remaining equations, one can verify that $v$ is constant, and
 \begin{equation}
 \ba{llll}
 a_{z}=\pm i \partial_{z}\ln f_1^* \,, ~&~ b_z=0\,,~&~ a_{t}=0 \,,~&~b_{t}=\pm 2 |f_1|^2\,.
 \ea
 \end{equation}
Interestingly enough, now the Gauss's law Eq.\eqref{gausslaw} gives, for $e^Y=|f_1|/v$,
 \be
 \partial_{z}\partial_{\bar z} Y-v^{4}e^{2Y}=0\,.
 \label{liuville}
 \ee
 This is the Liouville equation, albeit with the wrong sign for the potential term. It is well known that the most general solution is expressed in terms of a holomorphic function: let us call it $h(z)$, then we have
 \be
 e^{Y} = \frac{1}{v^2} \left| \frac{\partial_z h}{1 - h \bar h}
 \right| .
 \label{liouvillesol}
 \ee

 As we choose a particular holomorphic function $h(z)$, the complex plane is divided into two regions, depending on whether $|h(z)|>1$ or $|h(z)|<1$. The divergence at $h(z)=1$, or the delta function-like source, would be the M2-brane orthogonally intersecting with the M2-brane originally described by the BL theory. The intersecting locus is a one-dimensional curve on the complex plane. For instance, if we have
 \be
 h(z) = 1 + \frac{z_0}{z}
 \ee
 then $Y$ is divergent as we approach the line
 \be
 \mbox{Re} (z/z_0) = -1/2 .
 \ee

Now let us turn to the case of $\alpha \neq0$. We again find $v$ is constant, and without losing generality we may assume $\alpha$ is real and positive. One may thus set
\be
f_1 = \sqrt{\alpha} \cos\Upsilon
, \quad
f_2 = \sqrt{\alpha} \sin\Upsilon .
\ee
Then one can show that the gauge fields are given as
\be
\ba{llll}
 a_{z}= \partial_{z}\Upsilon\,, ~&~ b_{z}=0\,,~&~ a_{t}=0\,,~&~ b_{t}=\alpha\, \sinh (2\mbox{Im} \Upsilon)\,.
 \ea
 \ee
 Now the Gauss's law leads to, for $\Phi=\mbox{Im}\Upsilon$,
\be
2 \partial_{z}\partial_{\bar z}\Phi-\alpha v^{2}\sinh(2\Phi)=0 ,
\label{sG}
\ee
which is the sinh-Gordon equation.

One can construct a number of nontrivial solutions to Eq.\eqref{sG} using the B\"acklund transform. Consider the following coupled first-order differential equation,
\bea
\partial_z \Phi &=& +\partial_z \tilde{\Phi} - \beta \sinh (\Phi+\tilde{\Phi}) ,
\\
\partial_{\bar{z}} \Phi &=& -\partial_{\bar{z}} \tilde{\Phi} - \frac{\alpha v^2}{\beta} \sinh (\Phi-\tilde{\Phi}) .
\eea
Then one can easily verify that, for an arbitrary constant $\beta$, if $\tilde{\Phi}$ is a solution to Eq.\eqref{sG} then $\Phi$ is also a solution. Let us try simply $\tilde{\Phi}=0$. The above equations are easily integrated, and we get the 'one-soliton' solution,
\be
\cosh\Phi= \coth ( \beta z +\frac{\alpha v^2}{\beta}\bar{z} + z_0 ) ,
\ee
where $z_0$ is an integration constant. Obviously $\Phi$ is divergent at one point on the complex plane, when the argument of the $\coth$ is zero. We interpret this singularity as a signal of the intersecting M2-brane.



\subsection{M5-branes and fuzzy funnels}

\subsubsection{Without M-waves}
\label{ff:nowave}
It is again advantageous to start with the explicit projector and the corresponding
equations for 1/2-BPS. The generalized Nahm equations are considered already in Eq.~(\ref{gen:nahm})
with projector:
\be
P\epsilon =0 : \quad P = 1 \pm \Gamma_{y1234}
\ee
Since the supersymmetry parameter $\epsilon_0$ is chiral in $\mathbb{R}^8$ we can equivalently
use
\be
P = 1 - \textstyle{\frac{1}{2}} \cT_{IJKL} \Gamma_{IJKL} \Gamma_y
:\quad \cT_{1234}=\cT_{5678}=\pm 1
\ee

We need to find the most general configuration satisfying
\be
(D_\mu X^I \Gamma^\mu \Gamma^I
- \textstyle{\frac{1}{6}} F_{IJK} \Gamma_{IJK} ) P \epsilon=0
\label{BPSrel}
\ee
Using the explicit form of $P$, one can easily verify
\be
F_{IJK}\Gamma_{IJK} P
\epsilon
=F_{IJK}
\left(
2
\Gamma_{IJK} + 3  \cT_{IJPQ} \Gamma_{PQK}
+\cT_{IJKL} \Gamma_L
\right)
\epsilon
\label{onproj}
\ee
As we expand Eq.~(\ref{BPSrel}) there are terms with different numbers of the gamma matrix products.
Since $\epsilon$ is arbitrary, the terms with four gamma matrices should cancel with each other:
and the same applies to terms with 2-form, and 0-form. We then have the following set of equations
\be
\cT_{IJKL} D_y X_L + F_{IJK} = 0 ,
\quad D_t X^I = D_x X^I = 0 .
\ee
In general the corresponding M-brane configuration is given as follows:
\begin{table}[htb]
\begin{center}
\begin{tabular}{cccccccccccc}
   M2: & t & x & y & - & - & - & - & - & - & - & - \\
   M5: & t & x & - & 1 & 2 & 3 & 4 & - & - & - & - \\
   M5: & t & x & - & - & - & - & - & 5 & 6 & 7 & 8 \\
\end{tabular}
\caption{M-theory configuration for M2-branes and M5-branes :1/4-BPS of
the total 11 dimensional supersymmetry}
\label{m2:m5:m5}
\end{center}
\end{table}

One can easily generalize to BPS equations with less supersymmetry. For 1/4-BPS, one imposes another
projection rule, e.g.
\be
\Gamma_{y1234} \epsilon
=
\Gamma_{y1256} \epsilon
=
\epsilon
\ee
Then in general the M-theory interpretation is given as in Table~\ref{m2:m5:m5:quarter}.
\begin{table}[htb]
\begin{center}
\begin{tabular}{cccccccccccc}
   M2: & t & x & y & - & - & - & - & - & - & - & - \\
   M5: & t & x & - & 1 & 2 & 3 & 4 & - & - & - & - \\
   M5: & t & x & - & - & - & - & - & 5 & 6 & 7 & 8 \\
   M5: & t & x & - & 1 & 2 & - & - & 5 & 6 & - & - \\
   M5: & t & x & - & - & - & 3 & 4 & - & - & 7 & 8 \\
\end{tabular}
\caption{M-theory configuration for M2-branes expanded into two sets of intersecting M5-branes}
\label{m2:m5:m5:quarter}
\end{center}
\end{table}

Finally the minimally supersymmetric case in this class can be realized with
\be
\Gamma_{y1234} \epsilon
=
\Gamma_{y1256} \epsilon
=
\Gamma_{y1278} \epsilon
=
\epsilon
\ee
Then in general the M-theory interpretation is given in Table~\ref{m2:m5:m5:eighth}.
\begin{table}[htb]
\begin{center}
\begin{tabular}{cccccccccccc}
   M2: & t & x & y & - & - & - & - & - & - & - & - \\
   M5: & t & x & - & 1 & 2 & 3 & 4 & - & - & - & - \\
   M5: & t & x & - & - & - & - & - & 5 & 6 & 7 & 8 \\
   M5: & t & x & - & 1 & 2 & - & - & 5 & 6 & - & - \\
   M5: & t & x & - & - & - & 3 & 4 & - & - & 7 & 8 \\
   M5: & t & x & - & 1 & 2 & - & - & - & - & 7 & 8 \\
   M5: & t & x & - & - & - & 3 & 4 & 5 & 6 & - & - \\
\end{tabular}
\caption{M-theory configuration for M2-branes expanded into two sets of intersecting M5-branes}
\label{m2:m5:m5:eighth}
\end{center}
\end{table}

\subsubsection{With M-waves}
Starting from the 1/2-BPS fuzzy funnel equation, one can break the supersymmetry by introducing
a M-theory wave on the worldvolume of M2-branes.
The gravity wave along $x$ direction imposes an extra projection
\be
\Gamma_{tx}\epsilon=\epsilon
\ee
and the corresponding configuration is now as follows:
\begin{table}[h!]
\begin{center}
\begin{tabular}{lccccccccccc}
  MW: & t & x & - & - & - & - & - & - & - & - & - \\
   M2: & t & x & y & - & - & - & - & - & - & - & - \\
   M5: & t & x & - & 1 & 2 & 3 & 4 & - & - & - & - \\
   M5: & t & x & - & - & - & - & - & 5 & 6 & 7 & 8 \\
\end{tabular}
\caption{M-theory configuration for MW-M2-M2-M5}
\label{mw:m2:m5:m5}
\end{center}
\end{table}

And this can be repeated to any BPS configurations in Section~\ref{ff:nowave} and break the
supersymmetry into the half. So this class of solutions are rather exceptional in the sense that the
number of preserved supersymmetries can be odd. It is particularly intriguing that there exists
a set of BPS conditions with only one supersymmetry, i.e. 1/32 in the sense of whole 11 dimensional
supersymmetry.
\be
(D_t  - D_x) X^I=0, \quad D_\mu X^I - \textstyle{\frac{1}{6}} \cC^{IJKL} F_{JKL} = 0
\ee
where $\cC$ is the octonionic structure constant in eight dimensions.

In this class it turns out that one can construct BPS conditions from $N=1$ up to $N=7$.
With $N$ supersymmetries the R-symmetry is broken from $SO(8)$ to $SO(N)$.
In order to present the BPS equations it turns out most convenient to employ a set of rotation
generators in $SO(N) \subset SO(8)$. For instance let us consider $N=7$.
Rotation matrices $\cT_{IJ}$ are in ${\bf 28}$ of $SO(8)$ and
the branching rule is
\be
{\bf 28} \ra {\bf 21} + {\bf 7}
\ee
and it is the representation in ${\bf 7}$ which gives the BPS equation, since it can give a singlet when
multiplied to ${\bf 7}$, the Killing spinors.
In \cite{bps1} we identified the seven generators $\cT^{(p)}_{IJ} (p=1,\cdots,7)$ of $SO(7)$,
Then the BPS equations are written rather elegantly
\be
\cT^{(p)}_{IJ} D_\mu X^J + \textstyle{\frac{1}{2}} F_{IJK} \cT^{(p)}_{JK} = 0
\ee

\subsection{Kaluza-Klein monopoles and other special holonomy manifolds}
Now let us turn to the third class.
This type of BPS equations have been called "$SO(1,2)$-invariant" in \cite{bps1}, because they generically do not have spacetime dependence. Indeed the expression for the associated supercharge density does not involve spacetime derivatives, but they only have 3-product terms. And the BPS projectors involve only the $SO(8)$ gamma matrices.

For 1/2-BPS, the projection condition can be set simply as
\be
P\epsilon=0 : \quad P = 1\pm\Gamma_{1234}
\label{bps3:projection}
\ee
up to coordinate transformations. With this projection rule, the simplest BPS equation
one can think of is
\be
F_{125}  \pm F_{345} = 0
\label{bps3}
\ee
with all other fields set to zero. As discussed in Sec.3, We know that an infinite dimensional
3-algebra space is needed to find explicit solutions. One can for instance consider a direct sum of
two Diff${}^\infty(\mathbb{R}^3)$, i.e. $\tilde{X}_i,\tilde{Y}_i,\tilde{Z}_i, (i=1,2)$ satisfying
\bea
\left[ \tilde{X}_i , \tilde{Y}_j , \tilde{Z}_k \right] &=&
\left\{
\begin{array}{cl}
1 ,& \mbox{if} \,\, i=j=k \cr 0 , & \mbox{otherwise}
\end{array}
\right.
\\
\left[ \tilde{X}_i , \tilde{X}_j , \tilde{Y}_k \right] &=& 0 ,\quad \mbox{etc.}
\eea
Then obviously the following configuration will satisfy Eq.~(\ref{bps3}).
\be
X^1 = \tilde{X}_1 , \quad
X^2 = \tilde{Y}_1 , \quad
X^3 = \tilde{X}_2 , \quad
X^4 = \tilde{Y}_2 , \quad
X^5 = \tilde{Z}_1\mp\tilde{Z}_2
\ee
Our interpretation of this solution is that the M2-branes are puffed up first into M5-branes,
and then eventually they compose a Kaluza-Klein monopole, in $X^1,\cdots, X^5$ hyperplane.

For the most general 1/2-BPS configuration related to Eq.~(\ref{bps3:projection}),
one can just make use of the computation given in Eq.~(\ref{onproj}) and what follows,
and drop the spacetime derivative terms.
\be
F_{IJK} + \textstyle{\frac{3}{2}} F_{[I}^{\;\; LM} \cT_{JK]PQ}^{} = 0
\ee

If we generalize, other BPS configurations are constructed as supersymmetric
intersections or bound states of the KK monopoles given above. Since M-theory Kaluza-Klein
monopoles are purely geometric configurations which break the supersymmetry into 1/2,
they provide examples of non-compact K3 space. It is thus reasonable to relate the bound
states of KK with less supersymmetries as non-compact special holonomy manifolds in general.
As it is well known there exists a hierarchy of 8 dimensional spaces with different
numbers of unbroken supersymmetry. For each class there are a set of invariant tensors which
can be constructed as bi-spinor tensors. The projectors, for instance Eq.~(\ref{bps3:projection})
can be obviously expressed in terms of the invariant tensors, in this case the sum of volume form for
 each Kaluza-Klein monopole.
One then proceeds to find the most general BPS equations compatible with the given supersymmetry
by simplifying $F_{IJK}\Gamma_{IJK}P\epsilon_0=0$. This task has been performed in \cite{bps1},
and here we give a table which summarizes the result.
\begin{table}[htb]
\begin{center}
\begin{tabular}{|c|c|c|c|}
  \hline
   SUSY & Geometry & Invariant tensors & BPS condition \\
   \hline\hline
  2 & Spin(7) & 4-form $\cC$ & $\cC_{IJKL}F_{JKL}=0$ \\
  4 & SU(4) & Complex structure $J$ & $F$ is primitive $(2,1)$-form \\
  6 & Hyper-K\"ahler & $J^{(p)} \;(p=1,2,3)$ & $F_{IJK}J^{(p)}_{JK}=0$ \\
  8 & K3$\times$K3 & $\cT$ & $F_{IJK} + \frac{3}{2} F_{[I}^{\;\; LM} \cT_{JK]PQ}^{} = 0 $ \\
  10 & $D_k$ orbifold & $\cT, J$ & $\cN=8$ cond. and $F_{IJK}J_{JK}=0$ \\
  12 & $A_k$ orbifold & $J^{(p)}, \tilde{J}^{(p)}\; (p=1,2,3)$ & $F_{IJK}J_{JK}^{(p)}=
  F_{IJK}\tilde{J}^{(p)}_{JK}=0$ \\
  \hline
\end{tabular}
\end{center}
\caption{Summary of BPS equations}
\end{table}

\section{Discussions}
In this paper we have continued the study of BPS configurations in the Bagger-Lambert theory.
By considering the central extensions of the superalgebra, we have identified the numerous BPS equations of the BL action as M-branes. In general, they are suspersymmetric superpositions of various M-branes, ranging M-waves, M2-branes, M5-branes, and Kaluza-Klein monopoles.

Among the different types of solutions, we find the vortex solutions most intriguing. Using a particularly simplifying ansatz, we managed to reduce the coupled differential equations into Liouville or
sinh-Gordon equations. These are integrable systems, and in principle one can construct infinitely many nontrivial solutions. We have only considered the simplest solutions, and it would be interesting to systematically work out more solutions and study their properties, especially for the sinh-Gordon equation. Of course eventually we wish to address the case of non-constant $\alpha$ in Eq.\eqref{alpha_def}.

It is rather disappointing that all the explicit solutions we have presented in this paper are {\it singular}. Of course it is not unexpected, since the BL theory is devoid of a dimensionfull constant which could set the regularization scale. On the other hand, one can witness the fuzzy 3-sphere is balanced by the flux, in the mass-deformed version of the Bagger-Lambert theory \cite{mass_deformation}.

It is also quite interesting that using the infinite dimensional 3-algebra the Bagger-Lambert theory
can incorporate special holonomy manifolds. It will be nice if we can
derive, for instance the metrics of non-compact special holonomy manifolds from soliton solutions
of the Bagger-Lambert action.
Perhaps it helps to recall that
for ordinary Lie algebra structure of area-preserving diffeomorphisms
and the relevant non-commutative geometry, one can find a
precise mapping between non-commutative Yang-Mills instantons and gravitational instanton, e.g.
Eguchi-Hanson metric \cite{Yang:2005nr}.

More recently in \cite{Aharony:2008ug}, Aharony et. al. constructed $\cN=6$ Chern-Simons-matter action
with $U(N)\times U(N)$ gauge group, and claimed they describe $N$ M2-branes on $\mathbb{C}^4/\mathbb{Z}_k$ orbifolds.
The gauge theory is weakly coupled for large $k$, where the gravity dual is given by $AdS_4\times CP^3$.
It will be also interesting to explore the supersymmetric solutions of the $\cN=6$ theory, and see what they correspond to on the gravity side.

\acknowledgments
N. Kim wishes to acknowledge the
hospitality of the theory group at DESY, Hamburg during the visit. This research
has been supported by the Science Research Center Program of the
Korea Science and Engineering Foundation through the Center for
Quantum Spacetime (CQUeST) of Sogang University with grant number
R11-2005-021. N. Kim is also partly supported by the Korea
Research Foundation Grant No. KRF-2007-331-C00072.

\end{document}